\documentclass[runningheads]{llncs}
\usepackage{graphicx}
\begin{document}
\title{The Role of Functional Programming in 
Management and Orchestration of Virtualized
Network Resources\thanks{Supported by ERASMUS+ project ``Focusing Education on Composability, Comprehensibility and Correctness of Working Software'', no. 2017-1-SK01-KA203-035402 and the research project ``Reliability and Safety in Complex Software Systems: From Empirical Principles towards Theoretical Models in View of Industrial Applications (RELYSOFT)'' no. IP-2019-04-4216 funded by the Croatian Science Foundation.}}
\subtitle{Part II. Network Evolution and Design Principles}
\titlerunning{Management and Orchestration of Virtualized Network Resources}
%
\author{Tihana Galinac Grbac\inst{1}\orcidID{0000-0002-4351-4082} \and
Nikola Domazet\inst{2}}
\authorrunning{T. Galinac Grbac \and N. Domazet}
%
\institute{University Juraj Dobrila of Pula,
Zagreba\v{c}ka 30, HR-52100 Pula, Croatia
\email{tihana.galinac@unipu.hr}
\and
Ericsson Nikola Tesla, Krapinska 45, HR-10000 Zagreb, Croatia
\email{nikola.domazet@ericsson.com}}
\maketitle              
\begin{abstract}
This is part II of the follow-up lecture notes of the lectures given by the authors at the  
\emph{Three ``CO'' (Composability, Comprehensibility, Correctness)} Winter School held in Ko\v{s}ice, Slovakia, in January 2018, and Summer School held in Budapest, Hungary, in June 2019. In this part we explain the recent network evolution and the concept of virtualization, focusing on the management and orchestration of virtualized network resources. Network Functions Virtualization (NFV) is a new paradigm for changing the way networks are built and operated. Decoupling software implementation from network resources through a
virtualization layer introduces a need for developing sets of NFV management and
orchestration (MANO) functions. We discuss how this new point of view is highly inspired by the functional programming concepts. We provide examples and exercises on Open Stack virtual technology, and also discuss the challenges and problems inspired by telecommunication industry. Focus is on Reliable operation of Management and Orchestration functions of Virtualized resources.

These notes provide an introduction to the subject, with the goal
of explaining the necesity for new knowledge and skills in area of network programming. We introduce students with main problems and the network design principles, methods and techniques used for their solution.   
The worked examples and exercises serve students
as the teaching material, from which they can
learn how to use functional programming to
effectively and efficiently coordinate 
management and orchestration 
functions in distributed complex systems
using NFV.

\keywords{Network Function Virtualization \and
Management and orchestration \and Complex
software systems \and OpenStack platform.}
\end{abstract}

\section{Introduction}
\label{sect:Intro}
This lecture part II belongs to lecture series on the role of functional programming in management and orchestration of virtualized network resources. In the previous lectures part I of the follow-up lecture notes of the lectures given by the authors at the  
\emph{Three ``CO'' (Composability, Comprehensibility, Correctness)} Winter School held in Ko\v{s}ice, Slovakia, in January 2018, we discuss the \textbf{system structure for complex systems and design principles}. We provided introduction to
the theory of complex software systems reflecting on examples from telecommunication network and carefully positioning the considered problems imposed by network evolution and continuous complexity increase. Furthermore, we discussed main system design principles proposed to cope with complexity such as modularity, abstraction, layering and hierarchy. Since these are very generic recommendations how to design such complex systems we further explain in detail main paradigms such as service orientation and virtualisation forcing implementation of such principles. Virtualization is a paradigm frequently used in management of complex software systems. It implies introduction of a new abstract layer, a virtual edition of system layer and its functions, which avoids introducing dependency between system layers.

Here, in this lecture we go one step further where we discuss \textbf{network evolution and design principles}. We introduce new concepts that are cornerstones for future network evolution and are based on virtualisation and service orientation. These are Network Functions Virtualization (NFV) and Software Defined Networking (SDN).  Network Functions Virtualization (NFV) decouples network function from physical network resources through a new virtualization layer \cite{NFVchallenges} thus avoiding dependencies among them. However, it introduces a need for developing sets of NFV management and orchestration functions (MANO). Further in this lecture, we describe new challenges arising from implementation point of view and show students how to use the 
programming techniques for coordination of
management and orchestration functions of 
virtualized network resources operating in 
distributed environments. 

The problems and challenges of coordination of 
management and orchestration  
functions are addressed using the OpenStack  platform \cite{OpenStack}. It is an open source cloud operating system which integrates a collection of software modules that are necessary to provide cloud computing layered model. 
Such technology is necessary in dealing with 
problems arising from the virtualization paradigm
in current networks, and the students understanding
solutions in OpenStack will be able to 
transfer their knowledge to other existing
technologies with the same or similar purpose.

These notes provide an introduction to the subject, with the goal
of explaining the problems and the principles,
methods and techniques used for their solution.   
The worked examples and exercises serve students
as the teaching material, from which they can
learn how use of functional programming may result in
effective and efficient coordination  
management and orchestration 
functions in distributed complex systems
using NFV.

The methods and techniques explained in these 
lecture notes, and applied to the problems 
of management and orchestration of network
virtualization, are already existing and we
claim no originality in that sense. The purpose
of these notes is to serve as a teaching 
material for these methods. 

The challenges arising from the new network paradigms,
as well as their solutions, are illustrated 
through practical examples using OpenStack virtual 
technology and inspired by the problems from the
telecommunication industry.



The course is divided into following main parts:
 \begin{itemize}
    \item Background with reflection to key learnings from previous lectures on definition of complex system and challenging aspects of their management, system design principles and technologies enforcing design principles..
    \item New network technologies which drives network evolution such as Cloud Computing, Network Function Virtualisation and Software Defined Network.
    \item Management and orchestration of virtualized resurces and network design principles.
    \item Introduction to Open stack platform
    \item Reflections on practical examples.
\end{itemize}

 The main learning outcomes of this lectures 
 are to introduce virtualisation as one of the design principle for building modern complex systems, to explain the need of automated management and orchestration (MANO) functions in virtualized environments, to understand challenges of unreliable MANO functions in virtualized environments, and finally, to understand how well formalized virtualisation may help to improve reliable operation in network environments.










\section{Background}
\label{sect:Background}

Nowadays, all software systems and, more precisely, everything is getting interconnected over the Internet based telecommunication network. This network is distributed interconnecting various peripheral systems at the edge of the network, interconnecting variety of application domains.
Number of edge systems and its applications is increasingly growing thus forcing current core network to increase their capacities. Current networks are already getting very complex and their management becomes extremely expensive and inefficient. Therefore, new innovations are needed that would enable simplification of network management and use. 
System reliability and safety are of ultimate importance for ever growing range of applications and services. Note that in telecommunication network the services are provided to its users by distributed and complex systems in coordination. \textbf{Reliability} is defined as continuity of system functionality and service. \textbf{Safety} is defined as non-occurrence of catastrophic consequences on environments due to system unreliable operation.

The main problem of current research and practice is that we do not have adequate mathematical models that provide better understanding of underlying causes of such a complex system behavior and that can model global system properties that generate reliable and safe behaviour of modern software systems with increasingly growing complexity \cite{vanderMei2018}.  Network and system engineering principles have to be redesigned to accommodate these innovations. Current Software and Systems Engineering knowledge base has to be revised with new challenges \cite{Preeti2019,Denning}.
Furthermore, leading software industries (e.g. Google) have recognized these properties as vital specialization of \textbf{software and systems engineering} research that focuses on reliability and maintainability of large complex software systems and networks \cite{Beyer,Treynor}. This knowledge is recognized as important to next generation of software and system engineers with specialisation in network programmer. Hence, the setting of these lectures is within the theory of complex systems, in particular, the complex software systems and their role within telecommunication networks.

In aim of building an complex system there are numerous possibilities how to structure the complex system. The way how system is built is limiting or enabling its further evolution and system maintenance. Furthermore, in building large scale complex systems that provides complex functionalities the functional system composition is enforced as logical solution. This is especially the case with complex software systems present in the telecommunication network which is continuously evolving introducing more and more complex system functionalities, and whose whole evolution is following precise standards and recommendations described and regulated by numerous standard bodies. In fact, all these standards and recommendations define system functionalities which are achieved by implementing number of system functions. So, the functional system decomposition is already driven by the numerous standard bodies.

We provided introduction to the topic already in the first part of this lecture notes \emph{Part I. System structure for complex systems and design principles}  that we provided as follow--up lecture notes of the lectures given by the authors at the  
\emph{Three ``CO'' (Composability, Comprehensibility, Correctness)} Winter School held in Ko\v{s}ice, Slovakia, in January 2018, and Summer School held in Budapest, Hungary, in June 2019. Therefore, in the sequel we will just shortly recap the main learning and basic understanding that are needed to easy follow and understand advanced topics provided further in this lecture.

In the previous lecture, firstly, we started with an relevant definition of complex system from complex system theory, \cite{barabasi:Network-science} and apply this definition to complex \textit{software} system. \textbf{The complex software system} is a system where there exists a number of levels of abstraction and where it is impossible to derive simple rules from local system properties that are describing component behaviour towards global properties of system (such are for example reliability and safety). This behaviour of software systems is observed in the large scale systems like are mission critical systems that were evolutionary developed, which are usually very sensitive on reliability and security requirements. These systems are usually developed in sequence of projects and releases, involving several hundreds or even thousands of software and system engineers distributed around the globe, and product that is developed is exceeding several thousands lines of code that concurrently serves to millions of users in collaboration with similar complex systems in distributed network environment. There are many network nodes within the telecommunication network that share this challenges. In previous lecture we focus and interpret these challenges on mobile switching node that is central node for switching mobile subscribes within telecommunication core network.

Here, the main problem arise from the fact that human is developing these systems and as these systems grow the human inability to cope with such complexity is recognised as one of the main challenging obstacles to its further evolution. The main tool used to manage such software systems is system structure that is used to logically decompose complex system into set of system components that are needed to accomplish system functionalities. Such system structure is used to reason and manage system implementation while providing connection between local and global system properties, but more importantly to provide communication tool among all parties involved into development of such systems. Efficient systems use functional system decomposition which may serve to variety of system functionalities. In such system, side effects of changing system functions while implementing new and upgrading existing system functionalities we shall keep under control. Propagation of implementation effects or failures on variety of system functions may become very costly and time consuming. In this context, the functional programming paradigm is getting higher attention. The main idea behind is to treat program execution while operating system functionality as evaluation of mathematical functions without influencing global system state and keeping mutable data across system functions. However, this idea is not easy to achieve in such systems.

There are numerous possible candidate structures for building such systems and global system behaviour and system quality may be seriously influenced by selected candidate solution. To succeed as much as possible in the aim stated above we introduced the four main \textbf{system design principles}. These are modularity, abstraction, layering and hierarchy. \textbf{Modularity} means building systems as set of smaller system components that are independent of each other. \textbf{Abstraction} is term related to design of system interfaces and communications in between system components where the main idea is to design standard interfaces among components which are clearly separated from component internal implementation details. The components are further organized into \textbf{hierarchical layered} structure where components with similar functionality are grouped together within the system layer and communication follows strict hierarchical rules and only neighboured layers may communicate in between. In previous lecture we provide an overview of standard Open Systems Interconnection Model (OSI Model) which define hierarchical layering of system functions that are present in communication with other systems. Development of such standard have promote better interconnection within the equipment of different providers and across national borders and regulations. 

During the network evolution there is continuous grow in the number of possible network users, variety of technologies connected to the network and various services network offers. Core telecommunication network is continuously evolving finding new ways how to cope with new requirements such as massive traffic, with diverse information content, variety of different users, mobile and fixed, interconnected across geographic and application domains. The key technological trends implemented in modern telecommunication network are inspired by two main ideas, \textbf{virtualisation and service orientation}. 

These ideas are build within telecommunication network from the very beginning. Main motivation for virtualizing physical resources come along with first idea of building common telecommunication infrastructure that will provide its services to subscribers. This common infrastructure is shared among its subscribers. In previous lecture we provided detail description of introducing multiplexing number of subscribers within one common physical wire. The multiplexing of subscribers involved first abstraction of physical resource to its software representation. In order to implement reliable management over the shared resources proper virtualisation function has to be developed. The concept of service orientation has already implemented within the network. However, with network evolution the network service orientation is moving from manual process to software supported process. In modern telecommunication network, the user request services dynamically, whenever she or he needs the services, and network satisfies user needs by executing user request in fully service oriented computing paradigm. Even more, the network functions provide services one to another in service oriented fashion. 

Both of these concepts introduced numerous benefits such as increased capacity, enabling rapid innovation.

\section{Network evolution}
\label{sect:network-evolution}

Telecommunication networks are continuously evolved in generations and implements new concepts that enable to accomplish its main goal. The main goal during its evolution is to allow interconnection of various technologies by various vendors and in the same time to keep reasonable balance between costs and performances. Telecommunication networks are used by different classes of users, utilizing different technologies, sometimes with a very specific service demands. In such cases, a process of network configuration and management becomes very expensive and time and resource consuming. Efficient scaling of network resources, enabling innovation and introducing new services and energy efficient solutions are very hard to implement. The main problem network operators are facing today is how to effectively and efficiently manage high diversity of numerous users and technologies but at the same time achieve capital efficiency and flexibility for improvements.
Recent work is focused on development of new network architectures that would allow operators to architect its networks more efficiently. In sequel we introduce main ingredients of new network architecture defined for fifth generation (5G) network.

\subsection{Cloud Computing Platforms}
\label{sect:CloudComputingPlatforms}

\begin{figure}
\includegraphics[scale=0.75]{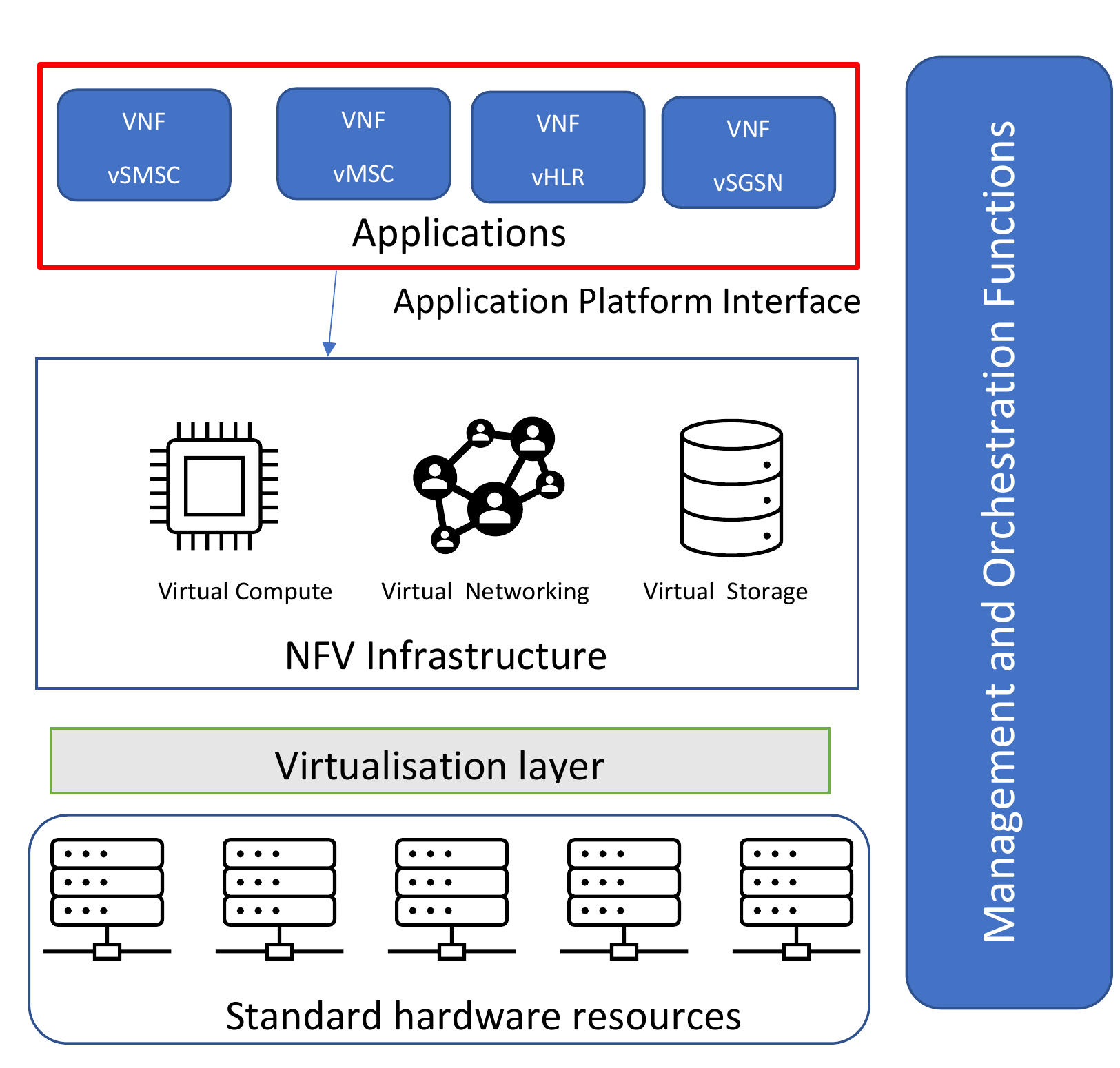}
\caption{Open stack virtualisation of network resources} \label{Openstackvirtualisation}
\end{figure}

There is growing need and interest in consuming computing and storage resources from third party vendors in as a service principle. For software development companies, the service orientation increase opportunities for specialisation while leaving hardware management operations out of its business. On the other side, vendor companies can specialize in hardware management business. Therefore, there is business driven need for open and well stanardized Application Platform Interfaces (API's) over which hardware vendors may offer its services to application service providers, see Figure \ref{Openstackvirtualisation}.

The new paradigm of abstracting resource plane requires huge efforts in standardisation of cloud platform. An operating system has to be developed for management of distributed hardware and related software resources and offering them as a services over the well standardised set of interfaces API's.  Note that this is key difference between distributed system and cloud system. Users may approach Cloud resources from single interface point (e.g. using command line interface or Graphical user interface) and use its resources on demand via well standardised API's. In traditional distributed system architectures all network resources were physical nodes with installed communication software for use on that single physical hardware. However, this paradigm has been changed and now communication software is independent from physical hardware and can be installed on any network node by using standard set of API's. This is the main reason why telecommunication systems are progressively moving into virtualized Cloud environments. 

With aim of speeding up this standardisation process of cloud platform there are established numerous initiatives. OpenStack is one such project established jointly by NASA and Rackspace intended to provide an open source cloud platform alternative that would be compatible with Amazon Elastic Compute Cloud (EC2). Furthermore, it should provide run time, reliable and massive scalability of resources with simple design. Therefore, to the project contributed numerous experts around the globe from various industries. Today, OpenStack becomes widely accepted as an innovation platform for Cloud platform industry \cite{radez-openstack-essentials,jackson:openstack-cloud-comp-cookbook}. Here, in this lecture all our examples will be provided on OpenStack with intention to provide examples of management functions and their operation in virtual environments. We selected an open source computing platform OpenStack aiming to simplify exercises execution to wide community, and especially targeting audience of graduate students at University Master level of Computing curricula.

\subsection{Network Function Virtalisation and Software Defined Network}
\label{sect:NFV_SDN}

\textbf{Network functions virtualisation (NFV)} term is referred to abstracting physical networking equipment and related behaviour by creating software representations (including memory and storage resources) of network elements and network operations. In other words, the NFV provides a network service that is decoupled from the physical hardware and offers feature set identical to and consistent to its hardware counterpart. Thus, network functions (hardware and software) are redesigned and offered as a service and following on demand principle and independently of the physical hardware. 
Network Functions Virtualisation (NFV) is aiming to define virtual network technologies that would allow operators to implement different technologies within its network offerings, for a which a dedicated and specialized device was needed by using common industry standard information technologies (IT), such as servers, switches and storage. 

The general framework arround implementation of NFV concept is defined in \cite{ETSIMANO} consist of following main layers:
\begin{itemize}
\item    Network functions virtualization infrastructure (NFVI) is the layer hosting generic COTS based hardware components like storage, compute, network hardware etc.
\item    Virtualized network functions (VNFs) is layer with functions implemented solely within software reusing benefits of software products like are easy scaling process, simple and fast deploying over multiple hardware, or even combining virtual instances on the same hardware, automation of these processes with licensing. 
\item Management and orchestration functions (MANO) that need to be developed for managing virtual instances and implementing its autonomous operation as we will discuss further withn this lecture. For this purpose a special working group is defined within the European Telecommunications Standards Institute (ETSI).  
\end{itemize}

\textbf{Software Defined Networking (SDN)} is a new networking paradigm which introduces additional abstractions in networks by separating a data and a control plane of networking devices. It assumes the control plane to be able to use standardized vertical interfaces to dynamically reconfigure the data plane flows, based on a global network policy. Therefore, many network functions can easily be virtualized using common servers and simple data plane networking devices.

Invention of Software Defined Network (SDN) architecture is motivated by the fact that traditional networking technologies are inadequate to scale to the levels required by today telecommunication networks. These limits are mainly caused by the complexity of network control and management functions and their distributed implementation logic. Distributed logic works well in medium sized networks but in today’s large and fast expanding network scenarios it becomes inefficient and too complex to manage and coordinate their scale and growth [15].  Therefore, a centralized network solution is needed. The main characteristics that should be provided by the solution are: 

\begin{itemize}
    \item Network management should be driven by general network objectives and low level device performance issues should be separated and considered at a lower level of abstraction.
    \item A global network view should be built and maintained for comprehensive understanding of network complexity at a higher abstraction level, such as its topology, traffic, and events.
    \item Devices at the lower level should be controllable through a standardized interface, such that they can be programmed and their behaviour changed on the fly, based on actual network demands and governed from the global network view.
\end{itemize}

The main fundamental basis of Software Defined Network are separation of Control and Data planes, simplified SDN devices (forwarding devices without complex distributed management protocols but managed from the control plane), centralized control (all network management is centralized at the control plane that is managing data plane SDN devices with help of an open standard), network automation and virtualisation and network openness. Open Networking Foundation (ONF) \cite{ONF} was established in 2011 by major network operators to promote adoption of SDN through open standards development. Open standards under consideration are Open Flow and Open Flow Configuration and Management Protocol, both used to communicate control decisions from the control to the data plane. The main idea behind SDN is to provide programmable network. The main challenges in SDN based networks are latency, scale, high availability and security. Latency may be affected with introduction of a central processing function. It may introduce delays because of numerous requests it has to process. Since number of users and devices connected to a network is continuously growing the question of scale in this centralised paradigm may be limited with processing power of the central function. Also, the central function has to be very reliable and highly available not to represent single point of failure for the whole network. Therefore, mechanism of high redundancy in processing and data storage may be required. And finally, central point of control may be serious issue for security attacks. 

\section{Management and orchestration of virtualized network resources}
\label{sect:Mgmt}

As is already stated in Sect.\ref{sect:Background} systems are getting more and more complex. The same situation is happening with the telecommunication networks. Networks are transforming from classical distributed set of interworking nodes to modern distributed interworking functions and services. The management of such complex system becomes very expensive, asking for higher expertise and higher skilled personnel in network management and consequences of actions performed are unpredictable. Note that in modern networked complex systems the functions are implemented in different functional blocks, as part of different complex systems, and that we need new knowledge in order to accomplish reliable operation for management and orchestration functions operating in these
distributed environments. Therefore, one of the recognized strategies in evolving telecommunication network is way towards its autonomy and self--management. Recent research efforts are devoted to innovation in this field. There is need for effective mechanisms to automate network so it may automatically adapt their configurations to new traffic demands and to introduce network flexibility and autonomously adapt to new technologies and new vendor equipment. These research efforts are driven by idea of autonomic computing \cite{Sterritt}, and further involve research on autonomic communication, autonomic networks, autonomic network management and self--managed networks. The final level of system autonomy is the level at which the humans only specify business policies and objectives to govern the systems while self--management following these policies is left to the system. Self--management mainly means:

\begin{itemize}
    \item self--configuration
    \item self--healing
    \item self--optimisation
    \item self--protection
\end{itemize}
In such new networks, the concept of programming software localized within one node, industry closed standard and solution for network functions is moved to concept of programming software for open network functions. The necessity for new profession of network developer is evident. In that new world of network programming we start to develop network design principles. In next section we open discussion on that topic. 

\subsection{Design principles for implementing autonomic behavior}
\label{subsect:autonomicDP}

Autonomic behaviour has been developed in many other fields and some general design principles have been recognized across all fields. In respect to network heterogenity, scalability and distribution the same principles may be valid also for networks. Here we shortly introduce these principles from \cite{AutonomicDesignPrinc} to motivate students to think about their implementation within examples provided in \ref{sect:Examples} of this lecture. 

\begin{itemize}
    \item Living systems inspired design
    \item Policy based design
    \item Context awareness design
    \item Self--similarity
    \item Adaptive design
    \item Knowledge based design
\end{itemize}

\textbf{Living system} inspired design is perspective to system design where inspiration is taken from functioning of living systems. There are many self--management mechanisms in functioning of living systems and in their interaction with environment and those ideas are taken as motivators for autonomy design. These concepts are mostly driven by survival instinct and collective behaviour. Survival instinct is related to system tension to come back to original equilibrium state. Collective behaviour refer to some spontaneous system reactions that may be derived from collective movement. Note that there is huge knowledge base derived from observing individual in respect to collective behaviour (like for example in Twiter, Facebook applications) and sometimes it happens that individual information moves collective behaviour in some particular state.

\textbf{Policy based} design is a predefined rule that governs behaviour of the system. This design principle have already been implemented widely across the network. However, it does not eliminate human interaction with the system.

\textbf{Context awareness} design is related to ability of the system to characterize situation or environment and based on historic behaviour decide how to adapt to new conditions. This principle have already been implemented within computing and networking field. One example are numerous sensing environment case studies.

\textbf{Self--similarity} design principle is related to characteristic that system organization persists as system scales and thus guarantee its global properties. This characteristic is also reflecting to global system properties that emerges solely of low level interactions, so low level interactions are not interfered with global. 

\textbf{Adaptive} design is related to ability of the system to adapt its inner behaviour as reaction to various environmental conditions. Such system is able to learn from its experience in operation and react accordingly by adapting its actions based on collected information and knowledge gained.   

\textbf{Knowledge--based} design is related to find the best design of knowledge gathering process. Since systems are complex, there are numerous possibilities in selecting appropriate set of properties to measure, designing appropriate data collection procedures and using appropriate artificial intelligence models to build appropriate knowledge base. This design is linked to building of appropriate business goals. 

\subsection{Current state}
\label{subsect:currentstate}

Networks are already highly developed and introduction of automation (by excluding human) into network management is not an easy and one step process. First step in automation is to virtualise its complex infrastructure and provide virtualized network resources. Furthermore, a real--time management and orchestration functions have to be developed that operate on these virtual network resources. As already mentioned, currently telecommunication network functions are progressively redesigned (to get virtual) so they can be offered over Cloud. In this process every network resource gets its own virtual image so it may be reinstalled, activated or deactivated as is needed in network reconfiguration or scaling demands. To automate these installation processes of this complex infrastructure a scripts are written which are then called for execution whenever needed during dynamic network management activities. These scripts are written in classical programming languages like is for example Python. Note here that real--time management and orchestration of network functions should secure avoiding the overlapping of management and ochestration processes over the same physical network resource pool. Again, functional programming like approach here is of ultimate importance to secure reliable and safe network management and orchestration operations.

\section{OpenStack}

\begin{figure}
\includegraphics[scale=0.75]{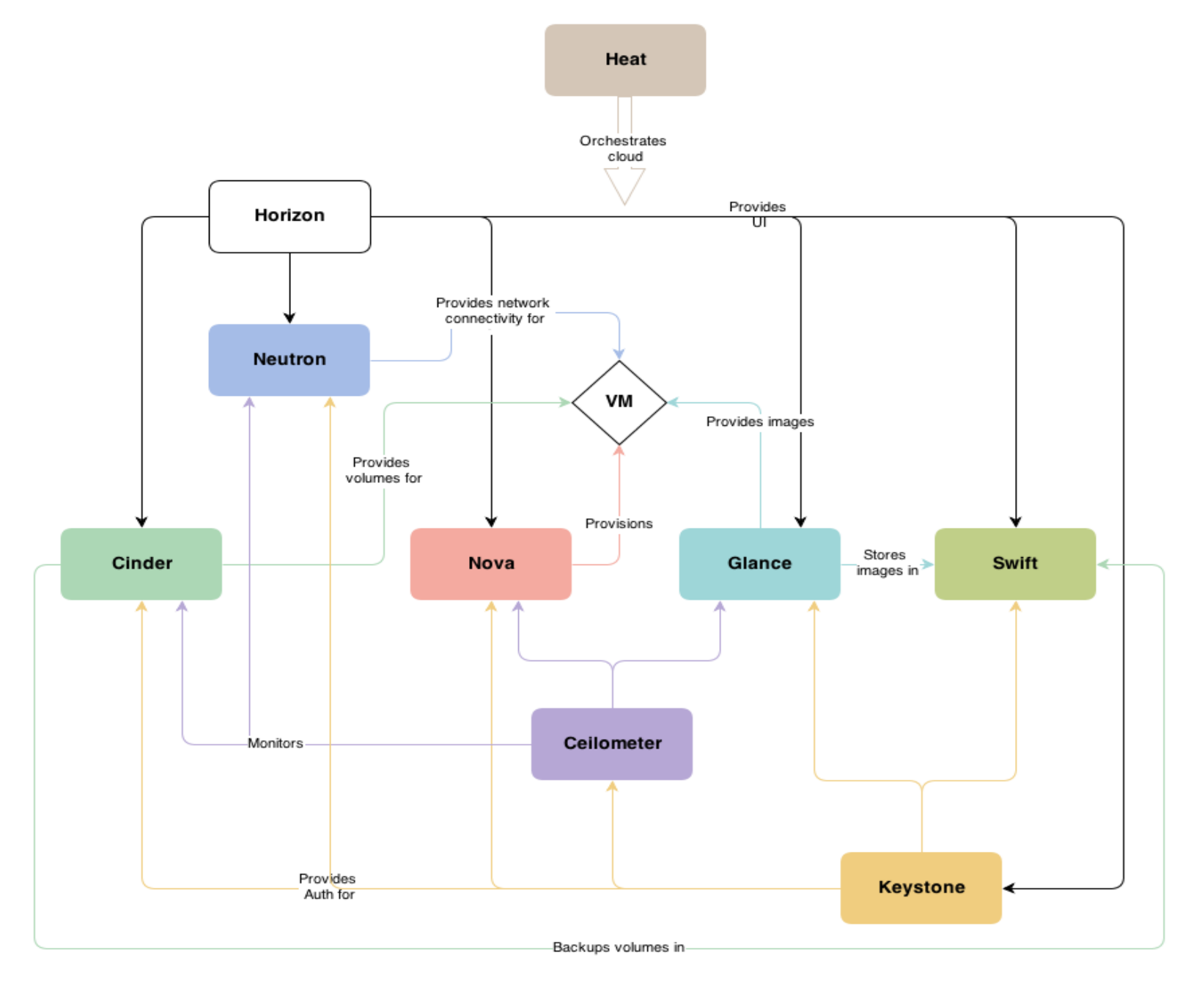}
\caption{Open stack conceptual architecture. Source www.openstack.org} \label{OSconcptarchitecture}
\end{figure}

OpenStack is a software platform that implements main functionality of providing distributed resources and infrastructure using 'As a service' paradigm to its users. 
Furthermore, OpenStack is a modular platform meaning that is designed as set of standardized units each designed to serve specific purpose, and these units may be used as needed or may be optional to OpenStack deployment. These units provide services to OpenStack users or other OpenStack units using standardised Application Platform Interfaces (API's). Table \ref{OpenStack_services} provides list of services, name of projects and short description of its main function.  The OpenStack was designed around three logical tiers: Network, Control and Compute, \cite{radez-openstack-essentials}. The Compute tier is taking over all the logic needed as hypervisor of virtual resources. For example, it implements agents and services to handle virtual machines. All communication among OpenStack services and with OpenStack users is provided through Application Platform Interface (API) services, web interface, database, and message bus. Numerous services have been implemented so far and detailed list of services can be found on OpenStack official web page and documentation \cite{OpenStack}. In aforementioned Table \ref{OpenStack_services} we listed just group of services specialized for specific purpose that we will also use in examples section \ref{sect:Examples} where we present how they operate together within an running Openstack environment. Furthermore, OpenStack offers communication through web interface called Horizon or dashboard. The Openstack conceptual architecture is presented in Figure \ref{OSconcptarchitecture} available from \cite{OpenStack} where is depicted interaction among OpenStack services mentioned in Table \ref{OpenStack_services}. 
For communication may be used MySQL, MariaDB and PostgreSQL databases and RabbitMQ, Opid, and ActiveMQ message buses.    

\begin{table}
\caption{OpenStack services and projects}\label{OpenStack_services}
\begin{tabular}{|l|l|p{7cm}|}
\hline
Projects & Services &  Short description\\
\hline
  Horizon & Dashboard & Web interface for using OpenStack services and manipulating with virtual resources.\\

Keystone & Identity service &  Authentification and authorisation functions. \\

Glance & Image service & Image Management services.  \\

Neutron & Networking service &  Provides services for networking of OpenStack resources to external network.\\

Nova & Compute service &   Lifecycle management of virtual resources.\\

Cinder & Block storage service &  Provides persistent storage functionality t virtual resources.\\

Swift & Object storage service & Data management over RESTful and HTTP based API's implementing fault tolerant mechanisms for data replication and scaling.\\

Ceilometer & Telemetry services & Collecting measurements and monitoring of resources\\

Heat & Orchestration service & Coordination of multiple virtual resources within one service provided to user\\ 

\hline
\end{tabular}
\end{table}

\subsection{Graphical User interface for manipulating virtual resources}
\label{subsect:dashboard}

\begin{figure}
\includegraphics[scale=0.75]{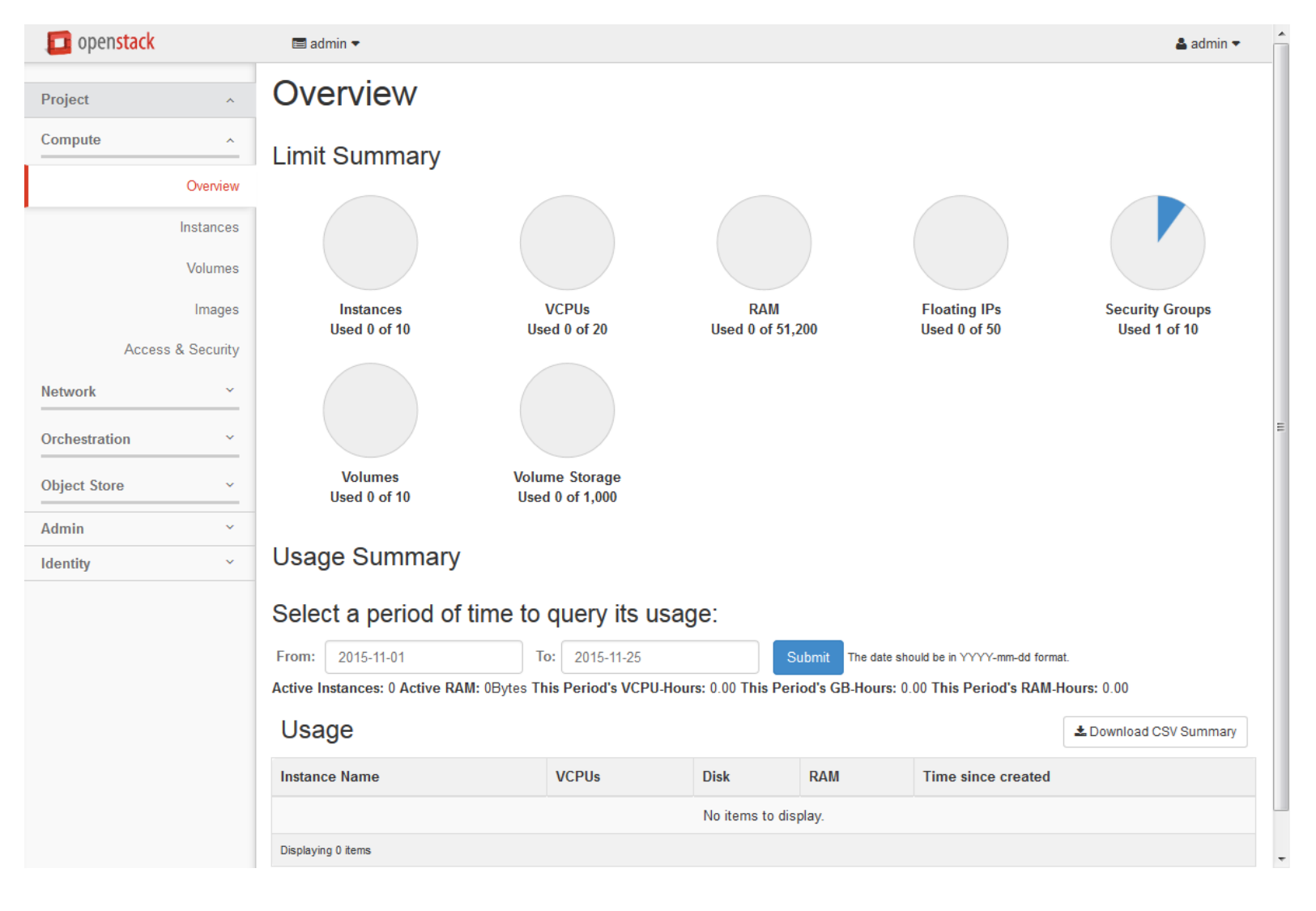}
\caption{Horison graphical user interface} \label{Horison}
\end{figure}

Horizon is a project defined within Openstack environment for management virtual resources over graphical user web interface.An screenshot of Horison GUI called dashboard is presented in Figure \ref{Horison}. Dashboard is a Openstack component that implements set of OpenStack services over the user interface. Actually, the OpenStack users are given the possibility to manipulate virtual resources over the visual commands provided on the web interface. In the background on the graphical user interface are implemented service calls to the API's of all officially supported services included within OpenStack. Note that OpenStack also provides a programmable access to its services over the API's that we describe in sequel. In the exercises we will more focus on programmable access.  

\subsection{Authentification and authorisation functions}
\label{subsect:keystone}

Authentication and authorisation of user access to cloud computing resources in OpenStack is managed through Keystone service. Objects that may be subject of keystone management operations are users, tenants, roles, instances (from catalog of services) and networks (endpoints of the virtual resources running in OpenStack environment).  

All objects must be assigned to tenants. Name tenant is used in command line while within dashboard the tenant is referred as project. A role has to be defined to each object assigned to tenant and its purpose is to restrict actions each object can perform. Even an administrator have to be defined its role and have to be assigned to tenant. Actions enabled for roles may be specified within a special policy documents, $/etc/PROJECT/policy.json$ files. 

Keystone maintains a service register or service catalog for the services offered by the components within the OpenStack. When a component is implemented within OpenStack cluster it should be registered in this service catalog. Service catalog contains a list of service names and related endpoints. The service endpoint is URL granted to this component within OpenStack cluster. The main benefit of this service catalog is that user only needs to know keystone address and the name of service which she or he wants to access. Then the keystone service is responsible to verify authentification of users and based on its role to verifiy if it is authorised to access the service. User never access Openstack services directly, it does always over the keystone service. Another important aspect of maintaining the service  catalog is in managing independency between users and local OpenStack implementation so the changes in endpoints are not propagated to all its users. I.e. this means that when an service changes its implementation and is deployed on another endpoint, the end user does not to be informed about that action. Service users will get correct service endpoint address by asking the keystone service just in time the service is needed.   

\subsection{Management of disk images}
\label{subsect:glance}

Glance is a component within Openstack with main function to manage disk images. For quick and fast deployment of virtual resources an pre--installed disk image may be used to boot from. Glance maintain the register of these disk images which are cached to compute node during instantiation of virtual resources and then copied to the ephemeral virtual resource disk location. These images had installed operating system but have removed secure identity elements such as Secure Shell host key (SSH) and network device MAC address that make this images generic and easily transferable to number of virtual machines without risk of interleaving the processes among them. These host specific information are transffered at system boot within a cloud--init script.

Disk images may be also made for specific purposes. For example if there is a multiple need for a specific web service, then the pre--instaled disk image may contain also web service preinstalation so the deployment process may be fully automated and faster for number of instances. There are available numerous tools for creation of such disk images with separated cloud-int script, like for example appliance-creator, Oz, and many others.

\subsection{Network management functions}
\label{subsect:neutron}
The main function of Neutron component is network management and offers to its users Networking as A Service (NaaS) functionality. This function is needed for configuring virtual resources to operate within virtual network environment. OpenStack uses Open vSwitch plugin to allow software defined networking of networking infrastructure and it provides a number of API's and related services for its management. These include, connection of virtual instances to virtual isolated networks, virtual routers, interconnection of virtual networks via virtual routers and to external networks via external gateways connected to virtual routers. Thus, users may configure its own virtual networks appliances which are interconnected to the external network. Neutron can manage multiple network appliances. 

Each instance may be associated to private or public network and is assigned private and public IP address range. Private or fixed IP address is assigned to an instance during its creation and is active during instance lifetime. On the other hand, an public IP address or floating IP address is not dependent of instance lifetime and it may be associated to an instance when the instance is made available for public and disassociated when instance is removed from public. Network Address Translation (NAT) transverse between public and private address spaces during communication flow between these two networks.

\subsection{Management of virtual instances}
\label{subsect:nova}

Nova is a component responsible for instance management. This includes managing of flavours, key pairs, instances, floating IPs and security groups. Flavors define amonut of resources that are alocated to an instance. Before an instance can be launched, authentification of users should be performed. An authenticated user use key pair (SSH pair) and security group to create its virtual instances. It can use its own SSH or the SSH generated by the system. The SSH key pairs are not new in OpenStack environment but it is reused principle from Linux. When an virtual instance is deployed a public key is placed in $authorized_keys$ file and running instance can be accessed using an SSH connection without password. Security group is a firewall at cloud infrastructure layer that should be opened to allow connection to virtual instance. By default, virtual instances belonging to the same security group may communicate to each other, while the rules should be specified for the Internet Control Message Protocol, SSH and other connections outside of the security group.

\subsection{Management of persistent memory}
\label{subsect:cinder}

Cinder is a component for management of block storage. It is used whenever a persistent memory space is needed, not dependent on instance lifetime. Note that disk space associated to an instance at its creation is destroyed at its termination. This is not the case for block storage. Block storage may be requested by users on demand and may be presented to running instance. It is also used for storing the snapshots of block volumes or of instances that are needed for instance boot.

\subsection{Management of object storage}
\label{subsect:swift}

Swift is object storage management component. In contrast to block storage, files and containers are stored without any metadata and are transferred from an instance to object store by using client--server communication with minimal overhead to the operating system. 

\subsection{Performance measurement functions}
\label{subsect:ceilometer}

An component within Openstack that is responsible for monitoring Openstack resources and collecting resource measurements is called Ceilometer. Originally it was designed for billing purposes but later it receives much generic purpose to take care about all telemetry within the OpenStack. These includes also observation of instance behaviour, its availability and performances, and for alarm setting. An very important application of ceilometer measurement system and alarm is for autoscaling of OpenStack resources at runtime.

\subsection{Orchestration functions}
\label{subsect:heat}

Openstack has special component responsible for orchastration of its resources. When multiple resources are intended to be used for the specific purpose and the same user these resources have to be interconnected and tied together so all operations that are available for regular Openstack instances may be also performed on this 'orchestrated' instance. For this purpose within a heat component of Openstack an template file may be used to specify resources that needs to be orchestarated, to specify their order and their mutual dependencies, required data that needs to be transferred among them. Heat is also compatible with Amazon Web Service (AWS) Cloud Formation template language.

\section{Examples}
\label{sect:Examples}

These exercises where developed for the purpose of Software Engineering Management course within Computer Science master study programme available on the following link $http://tania.unipu.hr/~tgalinac/OpenStack_Vjezbe-UPI.pdf$. The source files for the examples that follows could be accessed from the github $https://github.com/nikoladom91/CEFP2019$.

\subsection{Example 1}
\label{subsect:ex1}

Heat is the main project in the OpenStack Orchestration program. It allows deployment of resources on an OpenStack platform using templates. Heat supports various template formats and the format we will be using in this tutorial is the HOT (Heat Orchestration Template) format written as YAML files. 

The HOT files are executed by the Heat service and provide the blueprint for the deployment we want to achieve. A resource or groups of resources created during a HOT deployment is referred to as stack. We will use the following examples to describe the particulars of writing a HOT template and to show how ORCHESTRATION can be used.

\begin{example}
\label{ex1}

\texttt{\\ \\ heat\_template\_version: 2013-05-23}

\texttt{\\ description: Simple template to deploy a single compute instance}

\texttt{\\ resources:
\\ \indent my\_instance: 
\\ \indent\indent     type: OS::Nova::Server 
\\ \indent\indent     properties: 
\\ \indent\indent\indent       image: ubuntu\_cloud14 
\\ \indent\indent\indent      flavor: m1.small 
\\ \indent\indent\indent      key\_name: my\_key1 
\\ \indent\indent\indent      networks: 
\\ \indent\indent\indent\indent        - network: my\_net1 
}
\end{example}

In Example \ref{ex1} we use a basic template to explain the minimum required information for writing a functional template. We will go over the specific parts and describe their purpose.

The heat\_template\_version key is required in every template and it describes what version of HOT the template is written in. The description is optional and is usually used to describe the purpose and function of the template.

The resource section describes the resources the template will be creating and configuring during the deployment. It is required to have at least one resource per template. Each resource must have a type specified. This is used to deploy a specific OpenStack resource such as a virtual machine, nova network, security group etc. The list of available resource types for OpenStack version Mitaka can be found on the web-page \texttt{https://docs.openstack.org/heat/mitaka/template
\_guide/openstack.html}. The available resources somewhat differ between OpenStack versions so the correct one must be referenced when looking for them.

Services might require properties that contain the information required for their successful deployment. Some properties under the properties section are mandatory while others are optional. The properties for a resource are described under its type. Example 1 deploys a stack containing a single VM with hard-coded property values. The resource is identified as “my\_instance” and is of type “OS::Nova::Server”. Its properties describe what image and flavor will be used in the VM deployment, what security key will be provided to the OS and to what neutron network the vNIC of the VM will be connected. All the input resources used as properties need to be defined beforehand or the deployment of the stack will not be successful. Example 1 is not meant to be deployed, although it would deploy successfully. We will go over deploying a template after introducing Example 2.

\subsection{Example 2}
\label{subsect:ex2}

\begin{example}
\label{ex2}

\texttt{\\ \\ heat\_template\_version: 2013-05-23}

\texttt{\\ description: Simple template to deploy a single compute instance}

\texttt{
\\ parameters:
\\ \indent  image:
\\ \indent\indent    type: string
\\ \indent\indent    label: Image name or ID
\\ \indent\indent    description: Image to be used for compute instance
\\ \indent\indent    default: ubuntu\_cloud14
\\ \indent  flavor:
\\ \indent\indent    type: string
\\ \indent\indent    label: Flavor
\\ \indent\indent    description: Type of instance (flavor) to be used
\\ \indent\indent    default: m1.small
\\ \indent  key:
\\ \indent\indent    type: string
\\ \indent\indent    label: Key name
\\ \indent\indent    description: Name of key-pair to be used for compute instance
\\ \indent\indent    default: my\_key1
\\ \indent  private\_network:
\\ \indent\indent    type: string
\\ \indent\indent    label: Private network name or ID
\\ \indent\indent    description: Network to attach instance to.
\\ \indent\indent    default: my\_net1
}

\texttt{
\\ resources:
\\ \indent  my\_instance:
\\ \indent\indent    type: OS::Nova::Server
\\ \indent\indent    properties:
\\ \indent\indent\indent      image: \{ get\_param: image \}
\\ \indent\indent\indent      flavor: \{ get\_param: flavor \}
\\ \indent\indent\indent      key\_name: \{ get\_param: key \}
\\ \indent\indent\indent      networks:
\\ \indent\indent\indent\indent        - network: \{ get\_param: private\_network \}
\\ outputs:
\\ \indent  instance\_ip:
\\ \indent\indent    description: IP address of the instance
\\ value: \{ get\_attr: [my\_instance, first\_address] \}
}
\end{example}

To allow for the deployment of multiple stacks using the same template, input is needed. The optional parameters section is used to allow input. Unlike resources that represent an OpenStack resource entity, like a VM or a network, parameters represent certain values that are passed to the stack on deployment. Specific parameters are named, similar to specific resources, and are described by attributes. The type attribute is the only mandatory attribute and it defines the type of the value that the parameter represents. The label and description attributes are human readable parameter name and description and the default attribute describes the value that the parameter takes if no other value is given. There are more optional attributes that are not covered in this example.

The resource property uses an input parameter with the syntax \\ 

\centerline{\texttt{ "<property name>: \{ get\_param: <parameter name> \}"}. } 

\bigskip

Upon deployment the resource property will assume the value of the specified parameter. This allows the user to deploy a HOT multiple times with different input parameters and create unique stacks. The stacks may share the same blueprint but are separate entities with potentially different functionalities.
The outputs section allows for specifying output parameters available to users once the template has been deployed. We will see its use in later examples. Here we use it to output the IP of the VM we created as the parameter instance\_ip. The resource attribute value is retrieved with the syntax 

\bigskip
\texttt{"\{ get\_pattr: [<resource name>, <attribute name>] \}" }. 
\bigskip

This is used to retrieve resource attributes generated during deployment that can be used as outputs of the stack or as inputs for other resources.
Example 2 deploys a stack similar to Example 1 but, unlike Example 1, it can be passed different values for its deployment. If no new values are given, the specified default values will be used and the stacks from Example 1 and Example 2 will functionally be the same. They will still be separate entities as different UUIDs (Universally Unique Identifier) will be generated for the created resources. Providing different input parameters, VMs with, amongst other things, different images can be created creating functionally different resources.

\begin{example}
\label{ex3}

\texttt{\\ \\ ... \\}
\texttt{ 
\\ resources:
\\ \indent  rng:
\\ \indent\indent    type: OS::Heat::RandomString
\\ \indent\indent    properties:
\\ \indent\indent\indent      length: 4
\\ \indent\indent\indent      sequence: digits
}

\texttt{
\\ \indent  inst\_simple:
\\ \indent\indent    type: OS::Nova::Server
\\ \indent\indent    properties:
\\ \\ ... \\
\\ \indent\indent\indent      user\_data\_format: RAW
\\ \indent\indent\indent      user\_data: |
\\ \indent\indent\indent\indent        \#!/bin/sh
\\ \indent\indent\indent\indent        echo "Hello, World!" >> hello.txt
}

\texttt{
\\ \indent  inst\_advanced:
\\ \indent\indent    type: OS::Nova::Server
\\ \indent\indent    properties:
\\ \\ ... \\
\\ \indent\indent\indent      user\_data\_format: RAW
\\ \indent\indent\indent      user\_data:
\\ \indent\indent\indent\indent        str\_replace:
\\ \indent\indent\indent\indent\indent          params:
\\ \indent\indent\indent\indent\indent\indent            \_\_name\_\_: \{ get\_param: name \}
\\ \indent\indent\indent\indent\indent\indent            \_\_rnum\_\_: { get\_attr: [rng, value] }
\\ \indent\indent\indent\indent\indent          template: |
\\ \indent\indent\indent\indent\indent\indent            \#!/bin/sh
\\ \indent\indent\indent\indent\indent\indent            echo "Hello, my name is \_\_name\_\_. Here is a random 
number: \_\_rnum\_\_." 
\\ >> hello.txt
}
\end{example}

To automate certain procedures, users can pass blobs of data that the VM can access trough the metadata service or config drive. VMs that employ services like cloud-init can use the data in various ways. The blob of data is defined in the resource property “user\_data”. If given without additional attributes, the value of user\_data will be passed. If given the params and template attributes, the targeted text string defined under params is, within the text under template, replaced with the defined value. Example 3 replaces the “\_\_name\_\_” string with the parameter name while and “\_\_rnum\_\_” replaces it with a randomly generated number.

Here we can see the implementation of the get\_attr method where a value of a different resource is used within another resource. In this case a resource that when deployed represents a randomly generated number is created. The value of that resource is than used as an input for the data blob passed to the VM.

Example 3 HOT when deployed will generate a random number and instantiate two VMs. If the image used to instantiate a VM has the cloud-init service, that VM will execute the shell commands given in the user data as the root user. The inst\_simple VM will generate a hello.txt file in the / directory containing the “Hello, World!” string. The inst\_advanced VM creates the same file with the difference that the string within it contains the parameter name given as a HOT input and a randomly generated number.

\subsection{Example 4}
\label{subsect:ex4}

HOT allows for usage of nested code. This is done by defining the resource type as a bath to a different HOT file. It can be given as the path on the local environment from where the heat command is issued, or as a http/https link to a .yaml page accessible online containing the relevant HOT. When a nested HOT resource is defined, the input parameters are passed to that HOT trough the resource properties. The output parameters of the nested HOT are accessible as the resource attributes in the parent HOT.

When executing more complicated deployments with custom codes given as user data, heat cannot natively know if the given code has been executed correctly. The VM is deployed and Heat continues deploying other resources. Whether or not the code in the user data was successfully executed or how long it took is not taken in to account. If other resources depend on the successful execution of the user data code, it is needed to implement a waiting mechanic.

Heat provides two resources for the waiting mechanic. The OS::Heat::WaitCondition and the OS::Heat::WaitConditionHandle type resources. The OS::Heat::WaitCondition resource defines the waiting conditions. In the timeout property it defines how long the execution will wait for the HOT to complete before it is declared as a failed execution. The count property defines how many times a confirmation signal is expected before the execution is considered as successful. The handle property needs a link to the OS::Heat::WaitConditionHandle resource. That link is given with the get\_resource method.

The OS::Heat::WaitConditionHandle type resource is used to register the confirmation signal sent from the execution. It does this by defining an address that when curled with the appropriate information registers a confirmation signal. This curl command is inserted in to the user data code at the point where we want the confirmation signal to be sent, there can me multiple signals sent, each of which goes towards satisfying the count condition in the OS::Heat::WaitConditionHandle type resource.

Example 4 depicts a HOT which deploys two interdependent VMs. The first VM is a MySQL server. It is automatically configured during its initialization and when deployed is fully functional. The second VM is a Wordpress server that uses the MySQL database as its backend. As the Wordpress server requires for the MySQL database to be accessible during its initialization, the MySQL server employs the waiting service. The Wordpress VM initialization is therefor not started before the MySQL resource is deployed, as it requires some of its output attributes as its input parameters.

Each VM is started within a standalone HOT file which are both used as nested templates within the Example 4 script.

\textit{Stack deployment}

A resource or groups of resources created during a HOT deployment is referred to as stack. Here we will be describing how to deploy a stack using the Example 2 template. The deployment can be done via the Horizon web GUI or over the command line interface. The bellow command executed in the CLI will deploy the template from example 2. The template is fetched from github and the input parameters are passed as key=value pairs under the –parameters argument.


To list all deployed Heat Stacks, the command stack-list can be used as shown below.

\texttt{\\ \noindent Once \\} 

Once we know the UUID of a specific stack we can see its status and details with the command stack show as shown below.

Heat stack-show \texttt{<stack UUID>} 

\section{Use Case: Virtualisation of Mobile Switching Centre}
\label{sect:Usecases}

There are huge industry efforts to virtualise network functions that were developed in an closed industry product fashion. Some of the network products are older than forty years and are still active nodes within the curent telecommunication netwrk. One example is Ericsson Mobile Switching Centre node that was used as an example in Part I of this lecture series.

Mobile Switching centre implements communications switching functions, such as call set-up, release, and routing. It also, performs other duties, including routing SMS messages, conference calls, fax, and service billing as well as interfacing with other networks, such as the public switched telephone network (PSTN). This network function was actively been developed during 2G/3G generation networks. More information about this switching function can be found at 3GPP standards website (www.3gpp.org). 

This product has large installed base and is still progresivelly used in many operator networks. Therefore, it is estimated that operators will use 2G/3G networks as fallbacks for a long time to come, so it was decided to virtualise MSC to make it more compatible with modern environments.

There are identified numerous benefits of virtualizing this function. For instance, the virtual appliance of MSC function may be faster deployed and redeployed and thus it can be sold more quickly as only SW is needed for deployment. Both the product and the environment are scalable. Capacity increase is very simple; capacity of the product is increased by allocating more resorces to the VMs or deploying additional VMs, and capacity increase of the infrastructure itself would require adding more servers to the data centar. 
From here it may be concluded that virtualisation enables multiple products to run on the same datacenter and thus allowing operator more freedom in resource management. On the other hand side, the same data center could be used for multiple products, network functions and other virtualised instances thus eliminating the need for  hardware dedicated to every application domain.

Despite numerous benefits that virtualisation of MSC network function may imply there are also numerous potential problems that may arise on the way. In the case of Ericsson MSC, the product is developed in evolutionary fashion for more than forty years and as such it grows in complexity. Product has numerous functions that enables its long living but these functions were implemented highly relying on hardware aiming to satisfy very high reliability and safety requirements. To implement such hardware independent behaviour product has to be redesigned. Since the product is very complex because of number of functions implemented this act would require lot of expertise and cost.
Another very important aspect to understand is that mobile switching function that serves in real time services such as telephone call has very high reliability requirements and is usually higher that is the case with standard resources that are getting virtualised. For securing reliable operation of such virtualized MSC require additional layer that would secure this requirement. Therefore, Ericsson started developing a new project, its own proprietary network function virtualisation infrastructure called Ericsson Cloud Execution Environment CEE \cite{EricssonCM}. The product is developed by taking OpenStack as a base where proprietary solutions are incorporated to increase service reliability of virtualized products run on it. In Ericsson MSC not only software switching function was hardware related but also this special purpose hardware is implemented with special requirement to be reliable. The reliablity of this special purpose hardware is also much higher that is the case with standard equipment. Therefore, additional solution is to create specific data center for virtual network function purposes with high demands on performances.
There are other open source ongoing initiatives to produce High Available OpenStack solution such as for example OPNFV Doctor, OpenStack Freezer and OpenStack Masakari. All these solutions work on monitor, detact and correct solutions. However, the implementation solution for above stated design principles has to be invented and deployed within these solutions.

\section{Discussion and Conclusion}
\label{sect:Conclusion}

From the very beginning the telecommunication network has been built with the main aim to divide management of network cables and switches into separate business which would provide connection services to its users. In its core the switching board and network cables have implemented the multiplexing idea. With the help of the switching board, the same user can be involved in a number of connections (i.e., calls from subscriber's perspective or processes from processor's perspective) in the time sharing principle. This main multiplexing principle has been widely applied in every resource which is consumed in the network. During the network evolution calls/processes are multiplexed over each network cable, over the processors in switching nodes, over the memory in shared memory devices, etc. In the ongoing evolution step the processes are multiplexed over the shared network resources (not node resources) and even network functions are considered as network resources that users share in the time sharing principle. 

The above mentioned multiplexing or time sharing of common resources and providing them as a service is implemented by adding new abstraction layers and new virtualisation layer that introduce need for the new management functions securing safe and reliable switching of users on these common resources. 

The speciality of switching or multiplexing functions is in their high requirement on fast and reliable management. Since common resources are shared among its users in time sharing principle, every lost time slot is directly causing inefficiency and money loss. On the other hand, the services provided for each user must be safe and reliable, so that the user does not sense other users using the same shared resource. 

In all these evolutionary steps, there were specific switching programming languages which were used. In the essence of the functional programming is the ability to have functions that would for the given input always generate the same output. Thus, these functions can be easily formally verified by using mathematical logic. This is especially important in complex systems that require high safety and reliable operation. Although in complex time sharing systems it may be difficult to achieve pure functional programs, any good programmer should strive to get these programs as functional as possible. 

In telecom world, there are plenty of programming languages present in the switching domain. During history these languages evolved, so that functional programming languages, such as Erlang, have also taken dominance in this area. From the system verification point of view, the testers are used to work on a sequence of execution statements to easily follow the program execution. However, in pure functional world the failures would be minimised by proper formal methods. Hence, in fault mining process travelling across huge state machines would be avoided. Therefore, in principle, the more functional our code is, the less verification efforts would be needed. 

As we have seen, the complex software tends to become even more complex. Many software products started without functional programming paradigm and have become so complex that it would be too expensive and almost impossible to redesign them in functional programming fashion. However, new developments, especially those in which new abstractions are added and old source code is easily separated from the new code, should aim to move as much as possible to the functional paradigm. As we can see, evolution is just adding new abstractions and new management functions responsible for managing these virtual resources and implementation of these abstractions would be easier with pure functional code.   

In these Part II lectures, as well as in Part I, we went through the network evolution from the design principles and technology perspective. In Part I we introduced the main definition of a complex system, discussed challenges of their management. We introduced generic design principles for structuring software systems, such as modularity, abstraction, layering and hierarchy, in order to achieve their easier management. Furthermore, we introduced service orientation and virtualisation technologies that are used as a tool for implementing these principles. At the end of Part I, we discussed the example of the case study reporting experiences in redesigning existing complex software product with these design principles.  

In this Part II, as a continuation of the previous lecture, we introduced new evolutionary changes that are currently implemented within the networks. These are Network Function Virtualisation and Software Defined Networks. The two new concepts could be viewed just as adding new virtualisation layers on network resources (hardware and software functions) and introduce more service orientation and computation for each above mentioned network resource. Therefore, in addition to design principles stated in the previous Part I lectures that are related to structuring of complex software, we introduced now in Part II the design principles for implementing network autonomic behaviour. For the purpose of introducing the students with new technological changes, we provide examples by implementing simple network applications over the OpenStack platform by assuring aforementioned design principles for autonomic behaviour. Furthermore, we discuss an example of implementing a complex software product as a network application capable to run over OpenStack platform. Along with the example, we discussed benefits and problems that may arise in such act. Finally, we conclude with reflections on the role of functional programming in such complex networked environments.


\begin{thebibliography}{10}
\providecommand{\url}[1]{\texttt{#1}}
\providecommand{\urlprefix}{URL }
\providecommand{\doi}[1]{https://doi.org/#1}

\bibitem{AutonomicDesignPrinc}
Agoulmine, N.: Autonomic Network Management Principles: From Concepts to
  Applications. Academic Press, Inc., USA, 1st edn. (2016)

\bibitem{barabasi:Network-science}
Barab\'{a}si, A.L.: Network science. Cambridge University Press, 1st edn.
  (2016)

\bibitem{Beyer}
Beyer, B., Jones, C., Petoff, J., Murphy, N.R.: Site Reliability Engineering:
  How Google Runs Production Systems. O'Reilly Media, Inc., 1st edn. (2016)

\bibitem{Denning}
Denning, P.J.: Software quality. Commun. {ACM}  \textbf{59}(9),  23--25 (2016).
  \doi{10.1145/2971327}, \url{https://doi.org/10.1145/2971327}

\bibitem{EricssonCM}
{Ericsson CM--HA}. Ericsson (2020),
  \url{http://cqr.committees.comsoc.org/files/2017/03/04-Kelly\_Krick.pdf},
  accessed Nov 11, 2020

\bibitem{ETSIMANO}
{ETSI Industry Specification Group (ISG) NFV}: {ETSI GS NFV-MAN 001 v1.1.1:
  Network Functions Virtualisation (NFV); Management and Orchestration}.
  European Telecommunications Standards Institute (ETSI) (2014),
  \url{https://www.etsi.org/deliver/etsi\_gs/NFV-MAN/001\_099/001/01.01.01\_60/gs\_NFV-MAN001v010101p.pdf},
  accessed July 1, 2018

\bibitem{vanderMei2018}
Ganchev, I., van~der Mei, R.D., van~den Berg, H. (eds.): State of the Art and
  Research Challenges in the Area of Autonomous Control for a Reliable Internet
  of Services, pp. 1--22. Springer International Publishing, Cham (2018).
  \doi{10.1007/978-3-319-90415-3\_1}

\bibitem{NFVchallenges}
Han, B., Gopalakrishnan, V., Ji, L., Lee, S.: Network function virtualization:
  Challenges and opportunities for innovations. {IEEE} Communications Magazine
  \textbf{53}(2),  90--97 (2015)

\bibitem{jackson:openstack-cloud-comp-cookbook}
Jackson, K.: OpenStack Cloud Computing Cookbook. Packt Publishing (2012)

\bibitem{Preeti2019}
Mangey~Ram, J.P.D. (ed.): Tools and Techniques in Software Reliability
  Modeling, pp. 281--295. Academic Press (2019)

\bibitem{ONF}
{Open Networking Foundation}. Open Networking Foundation (2018),
  \url{https://opennetworking.org/}, accessed July 1, 2018

\bibitem{OpenStack}
{OpenStack Cloud Software}. OpenStack Foundation (2018),
  \url{www.openstack.org}, accessed July 1, 2018

\bibitem{radez-openstack-essentials}
Radez, D.: OpenStack Essentials. Packt Publishing (2015)

\bibitem{Treynor}
Sloss, B.T., Nukala, S., Rau, V.: Metrics that matter. Commun. {ACM}
  \textbf{62}(4), ~88 (2019). \doi{10.1145/3303874},
  \url{https://doi.org/10.1145/3303874}

\bibitem{Sterritt}
Sterritt, R., Bustard, D.: Autonomic computing - a means of achieving
  dependability? In: 10th IEEE International Conference and Workshop on the
  Engineering of Computer-Based Systems, 2003. Proceedings. pp. 247--251 (2003)

\end{thebibliography}
\end{document}